# Advantages of optical modulation in terahertz imaging for study of graphene layers

R. Ivaškevičiūtė-Povilauskienė,[1,a)] A. Paddubskaya,[2] D. Seliuta,[1] D. Jokubauskis,[1] L. Minkevičius,[1] A. Urbanowicz,[1] I. Matulaitienė,[3] L. Mikoliūnaitė,[3] P. Kuzhir,[4] and G. Valušis[1]

**AFFILIATIONS**

[1]Department of Optoelectronics, Center for Physical Sciences and Technology, Sauletekio av. 3, Vilnius 10257, Lithuania
[2]Laboratory of Nanoelectromagnetics, Institute for Nuclear Problems of Belarusian State University, Bobruiskaya Str. 11, Minsk 220006, Belarus
[3]Department of Organic Chemistry, Center for Physical Sciences and Technology, Sauletekio av. 3, Vilnius 10257, Lithuania
[4]Department of Physics and Mathematics, Institute of Photonics, University of Eastern Finland, Yliopistokatu 7, Joensuu 80101, Finland

**Note:** This paper is part of the Special Topic on Microwave Absorption by Carbon-Based Materials and Structures.
[a)]Author to whom correspondence should be addressed: rusne.ivaskeviciute@ftmc.lt

## ABSTRACT

It was demonstrated that optical modulation together with simultaneous terahertz (THz) imaging application enables an increase in contrast by an order of magnitude, thereby illustrating the technique as a convenient contactless tool for characterization of graphene deposited on high-resistivity silicon substrates. It was shown that the single- and double-layer graphene can be discriminated and characterized via variation of THz image contrast using a discrete frequency in a continuous wave mode. Modulation depth of 45% has been reached, and the contrast variation from 0.16 up to 0.23 is exposed under laser illumination for the single- and double-layer graphene, respectively. The technique was applied in the development and investigation of graphene-based optical diffractive elements for THz imaging systems.

## I. INTRODUCTION

Rapid evolution of terahertz (THz) technology and imaging evolving into a broad range spectrum of applications[1] stimulates an intensive search for new solutions in design and fabrication of convenient-in-use imaging systems. As a rule, preferences are given to compact or portable systems, preferably free of any optical alignment and containing only room temperature operating compact emitters and receivers coupled together with flat optical elements. The latter can be fabricated, for instance, using metamaterials,[2,3] either high-resistivity silicon-based diffractive optics[4] or 3D printed optical components.[5]

Due to its exceptional optical properties, graphene can be assumed as one of the most promising materials for THz and infrared passive optical components fabrication, i.e., modulators, transistors, etc.,[6,7] where contactless tuning is strongly preferred. However, for this kind of applications, high quality of the material is strongly preferred; therefore, contactless characterization gains particular attention.

Since pristine graphene is nearly transparent for THz radiation, it needs to be doped through any external force, i.e., either electrically driven or optically excited, aiming to make it effectively operating as a functional optical element. As a rule, such an approach of graphene modulation in a transmission mode requires substrate transparent for THz radiation; i.e., undoped silicon and germanium can be a suitable choice.[8–10] More detailedly, these studies were concentrated on single-layer graphene placement and investigation on these substrates under illumination of relatively high laser power reaching 400 mW in the case of germanium[10] and 40 mW in the case of silicon.[8] One can note that the modulation depth was estimated to be of 94% and about 70%, respectively. In the development of compact systems, preferences are given to small powers for modulation; therefore, the modulation depth dependence on the laser power attains a special focus of exploration. It is worth noting that photomodulation depth vs the incident laser power was investigated, and the estimated values were found to be around 7% at 25 mW[9] and about 45% at the same power level.[8]

In this work, the potential of the graphene optical modulation technique was extended via its application in THz imaging, enabling thus characterization and discrimination of both the single- and double-layer graphene in a contactless way. It was shown that such a technique allows an increase in the contrast in an order of magnitude, indicating that it can serve as a convenient contactless tool for characterization of graphene deposited on high-resistivity silicon substrates. The optical excitation power was kept to 25 mW, and the modulation depth of 42% was reached for single-layer graphene and 45% for double-layer graphene. A strong increase in the THz image contrast, up to 0.16, for the single layer graphene and 0.23 for double-layer graphene under optical excitation demonstrates the ability of THz imaging in contactless characterization and discrimination of graphene layer quality. This technique can be found useful in fabrication of graphene-based diffractive optical elements for THz imaging systems and on-chip designs in integrated photonics.

## II. SAMPLE CHARACTERIZATION

To investigate optical modulation of graphene, two types of samples based on single-layer and double-layer graphene structures were fabricated on a high resistivity 460 $\mu$m thick Si substrate. Both samples—single- and double-layer graphene—were produced using commercially available CVD graphene ("Graphenea") grown on a copper substrate and covered with $\approx$60 nm thick PMMA. To place CVD graphene on an Si substrate, the "classical" wet transfer technique[11] was used. According to the standard procedure, after copper etching and washing with distilled water, the graphene/PMMA structure was transferred on an Si substrate and then annealed at 130 °C. To remove the PMMA layer, the chloroform solution was used. In order to fabricate the samples with a double-layer graphene structure, the same technique was repeated. A newly produced graphene/PMMA unit was deposited on top of the obtained graphene monolayer on the Si substrate. Then, the polymer was removed.

The quality of obtained samples was estimated by Raman spectroscopy. The measurements were carried out with a Renishaw inVia Raman spectrometer equipped with a thermoelectrically cooled (−70 °C) CCD camera. All the spectra were recorded using 532 nm laser excitation. To avoid laser-induced sample heating, the average power on the sample's surface was set below 2.3 mW. The OLYMPUS LCPlan N 50/0.65 NA objective was used to collect Raman spectra signals. The exposure time was 10 s, and each spectrum was collected per 10 scans, yielding total 100 s time accumulation. The Raman frequencies were calibrated using the polystyrene standard ASTM E 1840 spectrum. The relative intensities of the Raman signal (instrument response function) were calibrated by using luminescence of NIST Intensity Standard SRM 2241. For the imaging, the $xy$ piezo stage was used.

The Raman spectra are shown in Fig. 1(a). As one can see, characteristic peaks of graphene are clearly visible. For single-layer graphene, the D peak is located at 1337 cm$^{-1}$, the G peak at 1583 cm$^{-1}$, and the 2D peak at 2673 cm$^{-1}$. The D mode is caused by a disordered structure of graphene. In this case, the intensity ratio of $I(D)/I(G) = 0.3$. The small ratio of D and G peaks indicates high order in the system and thus good quality of graphene. This also correlates well with higher carrier mobility.[12] For double-layer graphene, the D peak is visible at 1346 cm$^{-1}$, the G peak at 1581 cm$^{-1}$, and the 2D peak at 2686 cm$^{-1}$. The D peak is also very weak $I(D)/I(G) = 0.2$, which implies on a good quality of graphene. The G band position, intensity, and its shape are very sensitive to the doping level and the mechanical strain, which can be related to the interaction with the Si substrate. Since the G and 2D bands depend differently on graphene doping,[13] the evaluation of the $I(2D)/I(G)$ ratio is not suitable in our case.

Addition of the second graphene layer increases full width at half maximum (FWHM) of G and 2D peaks. For single-layer graphene, the FWHM of the G peak is 19.2 cm$^{-1}$, while FWHM of the 2D peak is 29.0 cm$^{-1}$. For double-layer graphene, the FWHMs of G and 2D peaks are 21.6 cm$^{-1}$ and 40.9 cm$^{-1}$, respectively. As expected, adding of the second layer of graphene causes the 2D band splitting into the overlapping modes. It also causes the FWHM of 2D peak widening by 41%.[14] In addition, the central position of the 2D peak shifts to a higher wave number. After deposition of the second graphene layer, it has shifted by 13 cm$^{-1}$ as with the increase of the number of layers, the 2D peak shifts to a higher wave number.[15]

Furthermore, from the comparison of Raman spectra collected from the areas corresponding to the single and double graphene (the measurements have been done on one sample), it can be concluded that the intensity of the G mode increased up to two times after transferring of the second graphene layer on top of the first one (data are not presented). This result is in a good agreement with the principle that the Raman intensity of the G mode increases almost linearly with the increasing graphene layer number until approximately ten layers.[16] In both Raman spectra, D +D″ bands are observed at 2450 cm$^{-1}$. These lines are associated with the emission of two phonons in the structure.[17] These results confirm that both samples display good graphene quality, and they show that additional layer of graphene gives expected Raman spectral properties of monolayer graphene.

To check the quality of the whole graphene layers, the Raman mapping has been performed. The measurements were carried out with a WITec alpha300 R Raman spectrometer equipped with a thermoelectrically cooled (−60 °C) CCD camera. All the spectra were recorded using 532 nm laser excitation. The average power on the sample's surface was set around 6 mW. The ZEISS EC Epiplan Neofluar 100×/0.9 NA objective was used to collect Raman mapping signals. The integration time was 0.5 s, and 50 spectra per line were collected. The Raman frequencies were calibrated using the polystyrene spectrum. For the imaging, the $xy$ piezo stage was used.

Raman maps of the intensity of G and 2D modes for samples with single- and double-layer graphene, respectively, are shown in Figs. 1(c) and 1(d). A small variation in the intensity distribution of both modes and the absence of the D mode (data are not presented) indicate high uniformity of transferred CVD graphene. It was first demonstrated[18] that the effect of the mechanical strain and charge doping can be optically separated from each other by correlation analysis of the frequency position of G and 2D modes. The plot demonstrates the distribution of $\omega_G$ and $\omega_{2D}$ obtained across a single graphene layer, as shown in Fig. 1(b). Red and black dashed lines corresponding to the directions in which the variation due to doping and mechanical strain are considered to be constant as well as zero point ($1581.6 \pm 0.2$ cm$^{-1}$, $2676.9 \pm 0.7$ cm$^{-1}$) were added for clarity. As one can see, due to interaction with oxygen environment, the single graphene layer on an Si substrate is affected by hole doping. On the other hand, the interaction with an Si substrate can explain the observed mechanical strain of the graphene layer.

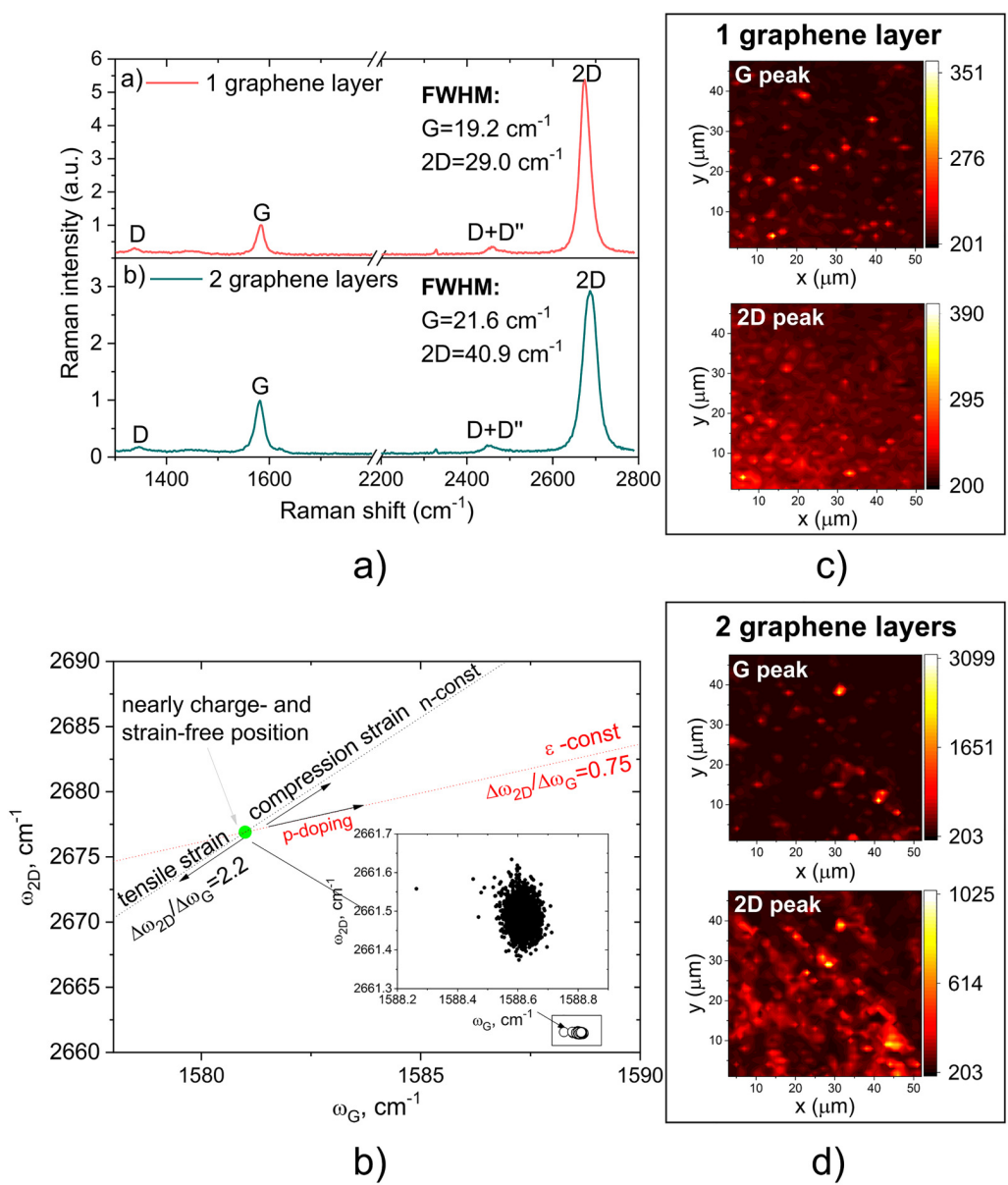

**FIG. 1.** (a) Raman spectra of single and double graphene layers. The intensity was normalized to G peaks for each sample. The spectra were excited by 532 nm laser limiting its power to 2.265 mW. (b) 2D and G peak frequency distribution for single-layer graphene. 2D and G peak mapping for (c) single-layer graphene and (d) double-layer graphene.

### III. TERAHERTZ EXPERIMENTAL SETUPS

Three different setups were employed in the experiments. The continuous wave (CW) frequency-domain THz spectrometer (Toptica TeraScan 780) in transmission geometry used to investigate properties of optically excited graphene in the THz frequency range is presented in Fig. 2(a). The emitted THz radiation is collimated by the first parabolic mirror (PM) and focused on the sample by the second PM. The beam spot diameter at the focal point of the second PM is 2 mm. Then, the transmitted beam again is collimated by the third PM and focused onto the detector by the fourth PM.

The THz mapping of graphene-based samples with and without optical excitation was recorded using a THz-CW imaging system [Fig. 2(b)], which provides additional information about the amplitude of the transmitted signal and its distribution along the sample surface. The THz-CW imaging system is based on a 0.3 THz electronic source (Virginia Diodes, VDI-175T) and includes the high-density polyethylene (HDPE) lenses and the antenna-coupled titanium microbolometer[19] as a THz detector with an aperture of 2 mm. The imaging was recorded by electronically modulating the source at 1 kHz frequency and detecting the microbolometer-induced signal by a lock-in amplifier.[20] Two crossed motorized translation stages were used to move samples in the x–y directions perpendicular to the incident THz beam, where the diameter in the focal point is 3 mm.

To evaluate the variation of the dynamic graphene conductivity, spectroscopic measurements were performed using a Teravil-Ekspla T-SPEC THz time-domain spectroscopic (THz-TDS) system based on a femtosecond laser (Toptica TeraScan, Femtofiber Pro). The system provides pulses of 780 nm wavelength, 90 fs pulse duration, and 150 mW output power with a 80 MHz repetition rate. Photoconductive antennas based on low-temperature grown GaAs were used for the emission and detection. The fast delay line was based on 10 times per second moving hollow retro-reflector with a 120 ps time window corresponding to 8 GHz spectral resolution. The diameter of the focused beam was 2 mm. The THz signal was detected by the digital signal processing card integrated into the electronic module with an analog-digital converter.

During all the experiments, the optical modulation was driven by a 666 nm continuous wave laser with a power of 25 mW and a beam spot diameter on the sample surface of 3 mm. The principal scheme is presented in Fig. 3(a).

## IV. RESULTS AND DISCUSSION

It is known that pristine graphene has a cone-like band structure, where its Fermi level is at Dirac's point [Fig. 3(b)].[21] Control of the electric conductivity can be done via tuning the Fermi level by doping graphene chemically, electrically, or optically.[22] Because of this cone-like band structure, density of states (DOS) of graphene is zero; therefore, when charge carriers are generated on the surface of graphene, Fermi energy rapidly increases or decreases depending on the p- or n-type of doping [Fig. 3(b)]. For smaller energies (IR or THz ranges), Fermi energy determines the optical conductivity of graphene.[23,24]

In this work, we used optical modulation induced by a pump laser. Since graphene's band structure is linear, optical absorption is independent of the excitation wavelength.

Transmission spectra of all samples were obtained using a THz spectroscopy system, and for each sample, it was measured twice—with and without laser illumination [Fig. 4(a)]. As one can see, transmittance varies due to the Fabry–Pérot interference and decreases under photoexcitation. Moreover, for samples with graphene, transmittance reduction is higher than that for pure Si. It is supposed that the optical injection of carriers generated by light absorption in Si is responsible for this effect. In the case of free-standing graphene, this modulation effect can be inverse because of the optical pumping, which increases the number of carriers and, consequently, the Drude absorption.[25]

The optical excitation was done by the pumping laser with $\lambda = 666$ nm, and the penetration depth $\delta_p$ in Si for this particular wavelength is 8 μm. Since the total thickness of Si was 460 μm, $\delta_p$ is relatively shallow and indicates that carriers generated under photoexcitation can easily be transferred to the graphene layers. Since graphene exhibits higher carrier mobility than Si, it causes an increase in electrical conductivity. According to (1) formula, higher conductivity results in reduced transmittance,[9]

$$\frac{T_{gr/Si}}{T_{Si}} = \frac{n+1}{n+1+Z_0\sigma(\omega)}; \quad (1)$$

here, $T_{gr/Si}$ is transmittance through graphene on an Si sample, $T_{Si}$ is transmittance through an Si substrate, $n = 3.42$ labels a refractive index of the substrate, $Z_0 = 377\,\Omega$ is impedance of free space, and $\sigma(\omega)$ denotes complex conductivity of graphene.

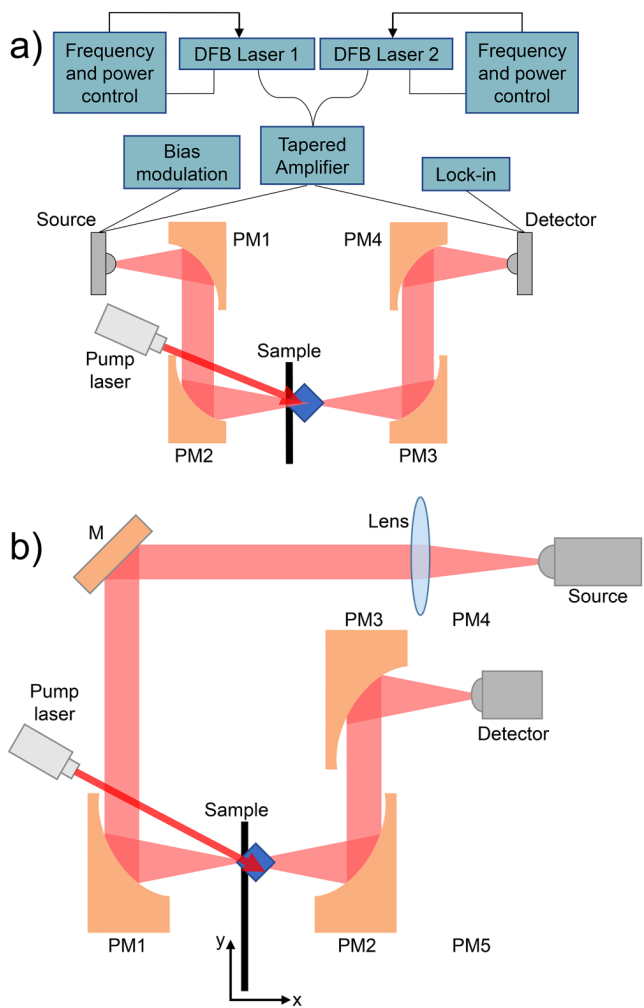

**FIG. 2.** (a) THz spectrometer setup (PM1–PM4—parabolic mirrors). (b) THz continuous wave THz imaging setup in transmission geometry (M, mirror; PM1–PM3, parabolic mirrors).

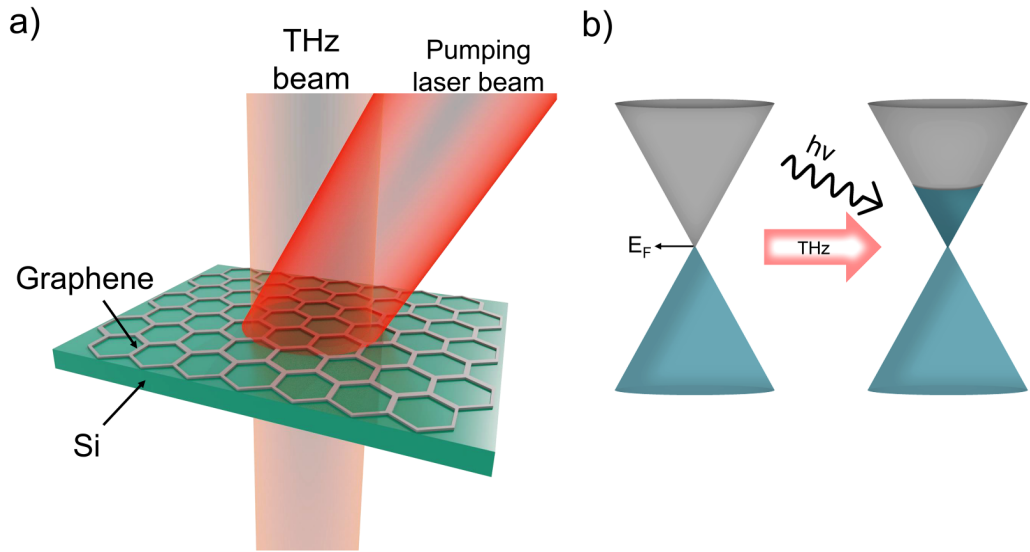

**FIG. 3.** (a) Principal scheme of the experiment. THz radiation and pumping laser beam overlapping at the sample surface. (b) Graphene band structure before and under laser illumination.

By adding the second layer of graphene, mobility of transferred free carriers increases, inducing thus a conductivity increase and, consequently, higher reduction in THz transmittance. Additionally, the modulation depth for all the samples was estimated by evaluating transmittance with and without photoexcitation using (2) and shown in the inset of Fig. 4(b),

$$M = \frac{T_{w/o} - T_{w/}}{T_{w/o}} \cdot 100\%; \quad (2)$$

here, $M$ is the modulation depth, $T_{w/o}$ labels transmittance without optical excitation, and $T_{w/}$ is transmittance with optical excitation.

As it was expected, the optical modulation is the weakest for pure Si and the maximum modulation depth is 14% at 0.43 THz frequency. Meanwhile, the maximum modulation depth reaches 42% for the single graphene layer and 45% for two graphene layers at 0.36 THz frequency.

Moreover, the modulation depth difference between single- and double-layer graphene is shown in Fig. 4(b). Shadows in the line indicate a 2% systematic error in the experiment. The maximum modulation depth difference between single and two graphene layers is well-resolved and reaches $12 \pm 2\%$ at 0.38 THz.

To evaluate a variation in electrical conductivity of graphene with and without optical excitation, THz time-domain spectroscopy (THz-TDS) measurements were carried out in two geometries—at the normal incidence of the THz beam and at the angle other than normal (as an example, at 43°). Such a rotation of the sample around the axis parallel to the component of the electric field of the incident THz beam is the simplest way to change pumping laser intensity (power/surface area) at the same point.

In contrast with the frequency-domain method, the THz-TDS technique is phase sensitive, which gives the possibility to evaluate the material parameters of the samples.[26,27] It was demonstrated[28] that for any thin conductive film deposited on a dielectric substrate, the transmittance can be defined using the following formula:

$$T_s = \frac{4 \cdot \alpha_s \cdot e^{idk_z} \cdot e^{-idk_z} \cdot e^{-idk_0}}{e^{idk_z} \cdot (1 + B_s - \alpha_s) \cdot (-1 + \alpha_s) + e^{-idk_z}(1 + \alpha_s) \cdot (1 + B_s + \alpha_s)}, \quad (3)$$

where for an s-polarized wave, $B_s = \sigma/\varepsilon_0 \cos[\theta]$, $k_0 = \omega/c$, $k_{0z} = k_0\cos[\theta]$, $k_z = k_0\sqrt{\varepsilon_s - (\sin[\theta])^2}$, and $\alpha_s = k_z/k_{0z}$. $\theta$ is the angle that is equal to 0° and 43° for considered cases.

Then electrical conductivity was calculated using Kubo's formula,

$$\sigma(\omega, \Gamma, \mu) = \frac{i \cdot 2e^2 k_B T}{\pi\hbar^2(\omega i\Gamma)} \cdot \log\left(2\cosh\left[\frac{\mu}{2k_B T}\right]\right)$$
$$+ \frac{e^2}{4\hbar}\left[\frac{1}{2} + \frac{1}{\pi}\arctan\left(\frac{\hbar\omega - 2\mu}{2k_B T}\right) \right.$$
$$\left. - \frac{i}{2\pi} \cdot \log\left(\frac{(\hbar\omega + 2\mu)^2}{\hbar\omega - 2\mu^2 + 4(k_B T)^2}\right)\right], \quad (4)$$

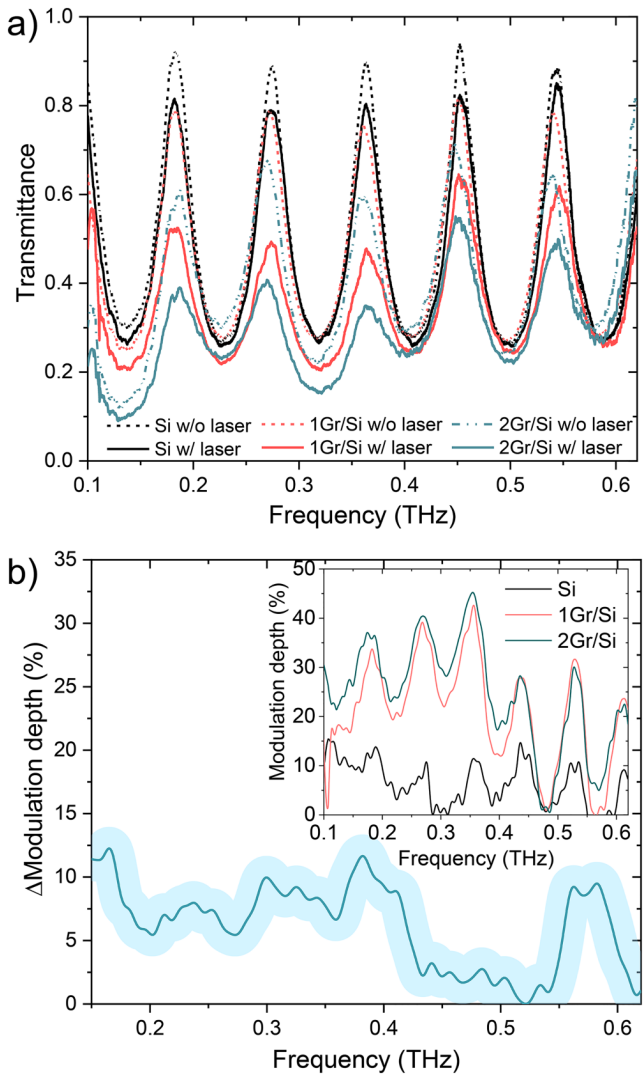

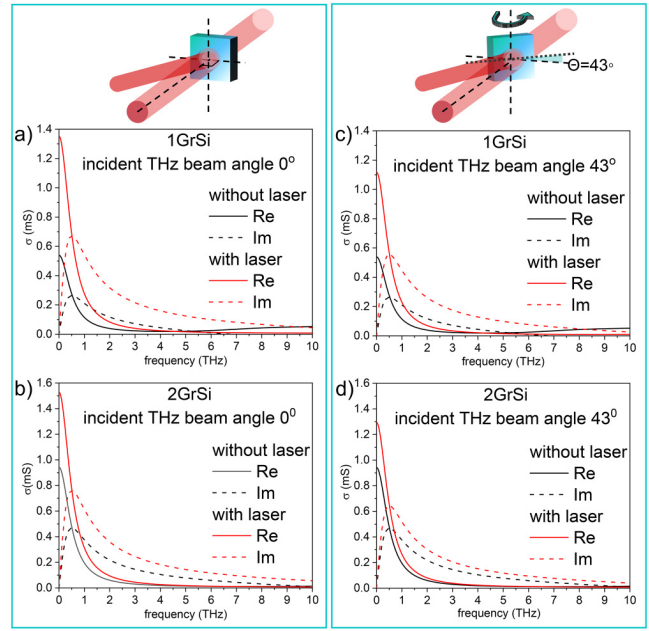

**FIG. 5.** *Re* and *Im* parts of (a) single- and (b) double-layer graphene conductivity when the incident THz radiation is perpendicular to the sample. *Re* and *Im* parts of (c) single-and (d) double-layer graphene conductivity when the THz radiation is at a 43° incident angle. The black line indicates results without photoexcitation and the red one with photoexcitation.

**FIG. 4.** (a) Transmittance spectra of the samples with and without photoexcitation. (b) Modulation depth difference between single- and double-layer graphene. Shadow of the line indicates a 2% systematic error in the experiment. The inset depicts the modulation depth of all investigated samples.

where $\omega$ is the angular frequency, $\Gamma = 1/\tau$, $\mu$ denotes the chemical potential, $k_B$ is the Boltzmann constant, $T = 293$ K depicts room temperature, and $\hbar$ indicates the reduced Planck constant.

It can be shown that formula (3) enables characterization of CVD graphene samples via parameter relaxation time $\tau = 50$ fs and chemical potential $\mu = 0.09$ eV. This technique allows one to evaluate the variation of the chemical potential of the single graphene layer from the value of 0.09–0.23 eV under photoexcitation at normal incidence and from 0.09 to 0.19 eV after sample rotation at 43°, which is in agreement with our predictions that the graphene doping level depends on the pumping laser intensity. The calculated *Re* and *Im* parts of the conductivity of the single graphene layer before and after excitation are presented in Figs. 5(a) and 5(c). In the case of the double graphene layers to have the best fit with the experimental data, it was necessary to take into account a weak interaction between the first and the second layers, which is in good agreement with our Raman results and a small difference in the doping level between the different graphene layers [Figs. 5(b) and 5(d)].

In order to evaluate the spatial distribution of transmittance changes under photoexcitation, THz imaging in transmission geometry was employed using the THz-CW imaging system operating at 0.3 THz. Results are presented in Fig. 6. Imaging results show that the single-layer graphene reduces THz transmittance by 5%, while the double-layer graphene reduces it up to 12%.

As it is illustrated in Fig. 4(a), the transmittance can be significantly modulated by photoexcitation. THz imaging results confirm that the transmittance change is minimal under photoexcitation for a pure Si substrate. However, the situation goes into sharp contrast when the single- and double-layer graphene is placed on the substrate and illuminated by the optical pumping laser. To qualitatively define the difference between the registered signal with and without laser illumination, contrast was defined as $C = |(T_{w/} - T_{w/o})/T_{w/}|$.[29] The contrast is 0.16 and 0.23 for single and double graphene layers, respectively. Meanwhile, the contrast value for pure Si is only 0.05.

These results point out that simultaneous application of a THz imaging technique with an optical excitation leads in well-resolved changes in the image contrast; hence, the effect can successfully be used for contactless graphene characterization and evaluation of the different number of graphene layers in the investigated structure.

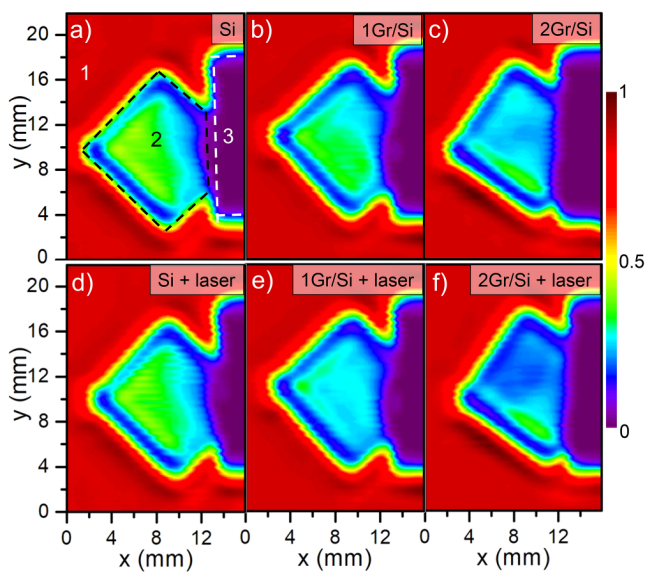

**FIG. 6.** 2D imaging of samples recorded at 0.3 THz in transmission geometry: without photoexcitation (a)–(c) and with photoexcitation (d)–(f). The transmittance scale is normalized to the maximum signal. Under photoexcitation, the resulting contrast is 0.16 for single-layer graphene and 0.23 for double-layer graphene. In (a), 1 marks the air, 2 the sample, and 3 the sample holder.

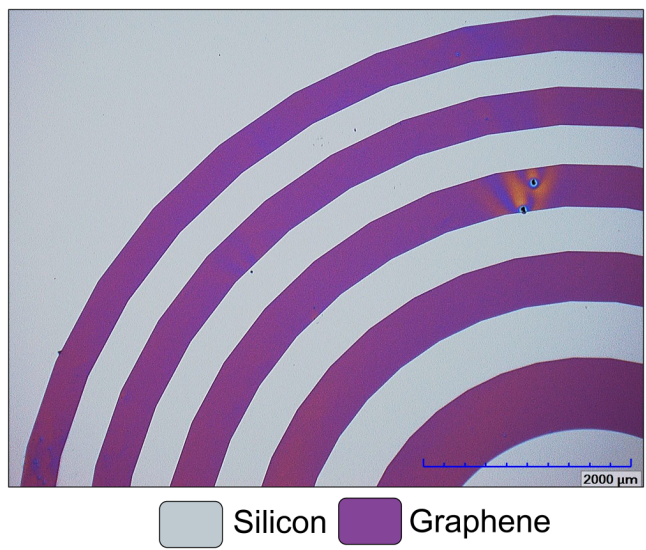

**FIG. 7.** Photo of the quarter of a graphene-based zone plate. The gray area indicates silicon, whereas the purple area depicts graphene parts.

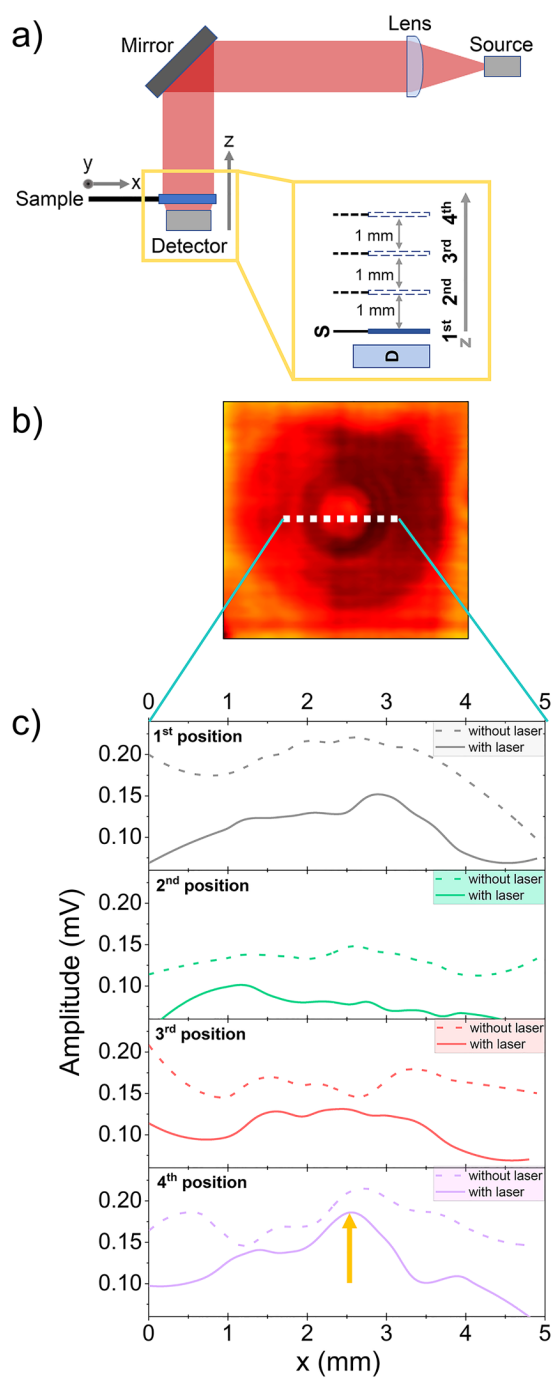

**FIG. 8.** (a) THz-CW system setup at 0.58 THz. The sample was shifted away from the detector, starting from the first position and moved further in a 1 mm step. (b) 2D imaging of investigated GZP. (c) Cross sections of transmitted THz radiation through GZP at first, second, third, and fourth positions indicated in panel (a). The evaluation is made from a 5 mm width part at the center depicted as a white dotted line in panel (b). The yellow arrow indicates pronounced focusing of the THz radiation.

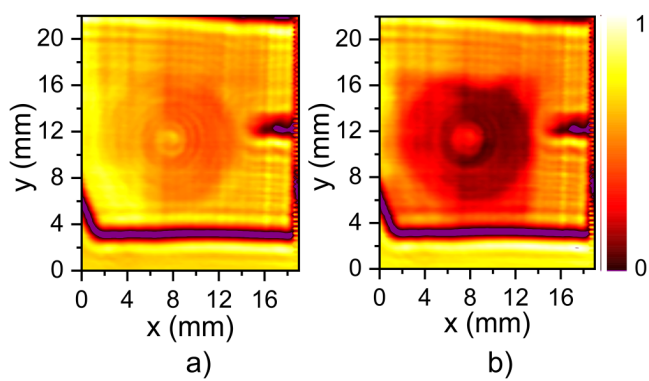

**FIG. 9.** Two-dimensional imaging of a graphene zone plate at 0.58 THz on an Si substrate (a) without and (b) with laser illumination. The colored scale indicates the transmitted signal normalized to the maximum value.

## V. GRAPHENE-BASED THz ZONE PLATE

THz zone plates as compact focusing elements can be useful in THz imaging systems with a broadband source when, e.g., for spectroscopic THz imaging aims, only one particularly selected frequency is needed to record the image.[20,30]

A graphene zone plate (GZP) (Fig. 7) was fabricated using a graphene patterning technique, where the single graphene layer was first transferred on the Si substrate in the same way as described in Sec. II. Then, the PMMA layer was coated on the top of graphene for protection. Subsequently, photolithography was performed after which patterned photoresist served as a mask for graphene. Later, step by step, each layer was removed until only patterned graphene in the shape of a zone plate was left on the Si substrate.

The zone plate was designed for 0.58 THz frequency; the diameter of the outer zone was 1 cm and the width of one stripe was ≈500 $\mu$m.

To examine GZP focusing abilities, THz imaging in the transmission geometry was recorded using a THz-CW system. The setup is shown in Fig. 8(a). Here, the same electronic THz source tuned to 0.58 THz as in Fig. 2(b) was used. The images were recorded moving GZP in $xy$ directions. Every full 2D scan was repeated by moving the GZP in the $z$ direction away from the detector, starting from the closest (first) position and then moving further in a 1 mm step.

Cross sections of the transmitted THz radiation were evaluated along the 5 mm line in the center of the beam, depicted as a white dotted line shown in Fig. 8(b). Cross sections at all four positions are shown in Fig. 8(c). As one can see at the first, second, and third positions, the mode cross section is smooth, about 4 mm wide with detection amplitudes of 0.15, 0.09, and 0.12 mV, respectively. At the fourth position, the profile exhibits peculiarities expressed as a higher peak with the reduced width to 1 mm and the increased amplitude to 0.19 mV (marked with a yellow arrow), indicating focused radiation at the center of the zone plate.

Two-dimensional imaging of a graphene zone plate at the fourth position with and without photoexcitation is displayed in Fig. 9. The colored scale indicates the amplitude of the transmitted signal normalized to the maximum value. As it is seen in Fig. 9(a), the transmitted signal is decreased in the zone patterned with graphene. The contrast at the center in the THz image without optical excitation is 0.44. Optical excitation changes the situation essentially in increasing the contrast up to 4.17 at 0.58 THz as it is presented in Fig. 9(b). Therefore, GZP can clearly be presented in THz imaging results. It implies that THz imaging combined simultaneously with optical modulation can serve as a convenient contactless tool for characterization and discrimination of graphene layers on silicon substrates.

## VI. CONCLUSIONS

Single- and double-layer graphene was deposited on the high-resistivity silicon substrates using a wet transfer technique. The measured Raman spectra indicated good quality of prepared layers. Optical modulation of single- and double-layer graphene is demonstrated with corresponding modulation depth values of 42% and 45%. THz time-domain spectroscopy enabled one to evaluate graphene conductivity before and under laser illumination, which exhibits an increase in its real part of ≈2.5 times for a single graphene layer and ≈1.6 times for a double graphene layer. Advantages of THz imaging with its simultaneous use of optical modulation were revealed. The pronounced change observed in contrast of 0.16 for single-layer graphene and 0.23 for double-layer graphene with and without laser illumination allowed one to infer that THz imaging can serve as a convenient contactless technique for characterization and discrimination of graphene layers on silicon substrates.

## ACKNOWLEDGMENTS

This work was supported by Horizon 2020 under Grant No. 823728 (DiSetCom). P.K. was supported by the Academy of Finland Flagship Programme "Photonics Research and Innovation (PREIN)" (Decision No. 320165).

## AUTHOR DECLARATIONS
### Conflict of Interest

The authors have no conflicts to disclose.

## DATA AVAILABILITY

The data that support the findings of this study are available from the corresponding author upon reasonable request.